\documentclass[twocolumn,usenatbib]{aa}
\usepackage{graphicx}
\usepackage{natbib}
\bibpunct{(}{)}{;}{a}{}{,} 

\usepackage{txfonts}
\begin{document}
   \title{The nuclear regions of NGC 7582 from [\ion{Ne}{II}] spectroscopy \\
   at 12.8$\mu$m -- an estimate of the black hole mass\thanks{Based on 
	observations obtained with VISIR at the ESO Very Large Telescope, 
	Paranal, Chile}}    

   \author{M. Wold \inst{1}, M. Lacy \inst{2}, H.U. K{\"a}ufl \inst{1},
	   R. Siebenmorgen \inst{1}}

   \offprints{M. Wold}

   \institute{European Southern Observatory, Karl-Schwarzschild str.\ 2,
              85748 Garching bei M{\"u}nchen, Germany
         \and
             Spitzer Science Center, California Institute of Technology,
             Caltech, MC 220-6, Pasadena, CA 91125, U.S.A. }

   \date{Received ; accepted }

\abstract
{
We present a high-resolution ($R\approx16,000$) spectrum and a narrow-band 
image centered on the [\ion{Ne}{ii}]12.8 $\mu$m line of the central kpc region
of the starburst/Seyfert~2 galaxy \object{NGC 7582}. The galaxy has a rotating 
circum-nuclear starburst disk, shown at great detail at a diffraction-limited 
resolution of 0\farcs4 ($\approx 40$ pc). The high spatial resolution allows 
us to probe the dynamics of the [\ion{Ne}{ii}] gas in the nuclear regions, and 
to estimate the mass of the central black hole. We construct models of gas 
disks rotating in the combined gravitational potential from the stellar bulge 
and a central black hole, and derive a black hole mass of 
$5.5 \times 10^{7}$ M$_{\odot}$ with a 95\% confidence interval of 
[3.6,8.1]$\times$10$^{7}$ M$_{\odot}$. The black hole mass combined with 
stellar velocity dispersion measurements from the literature shows that the 
galaxy is consistent with the local $M_{\rm BH}$-$\sigma_{*}$ relation. This 
is the first time that a black hole mass in a galaxy except our own Milky Way 
system has been estimated from gas dynamics in the mid-infrared. We show that 
spatially resolved mid-infrared spectroscopy may be competitive with similar 
techniques in the optical and near-infrared, and may prove to be important for 
estimating black hole masses in galaxies with strong nuclear dust obscuration.
The high spectral resolution allows us to determine the heliocentric systemic 
velocity of the galaxy to between 1614 and 1634 km\,s$^{-1}$. 
The mid-infrared image reveals several dense knots of dust-embedded
star formation in the circum-nuclear disk, and we briefly discuss its 
morphology.

   \keywords{galaxies:individual:NGC 7582 -- galaxies:nuclei -- 
             galaxies:starburst -- galaxies:Seyfert -- galaxies: star clusters
               }
   }
\authorrunning{Wold et al.}
\titlerunning{The nuclear regions of NGC 7582 from [\ion{Ne}{ii}]12.8$\mu$m 
 	spectroscopy}

   \maketitle

\section{Introduction}

Much effort is currently being put into understanding how black holes 
form and grow in the context of galaxy formation 
\citep[e.g.][]{kh00,marconi04,matteo05,cattaneo05}. The 
$M_{\rm BH}$--$\sigma_{*}$ 
relation \citep{fm00,gebhardt00}, being a tight correlation between black hole
mass and stellar velocity dispersion, is strong evidence for a causal link 
between the growth of the black hole and the galaxy bulge. 
The $M_{\rm BH}$--$\sigma_{*}$ relation has become our most 
important tool for studying the co-evolution of black holes and 
galaxies. To advance our knowledge we need to investigate how
black hole mass and bulge velocity dispersion relate in a variety of 
galaxies, over a wide range in Hubble type, black hole mass and velocity 
dispersion. However, direct measurements of black hole mass in galaxies 
are challenging because resolutions of the order of the radius of the 
sphere of influence of the black hole are required, 
a scale which is typically $\ll 1\arcsec$. 

A widely used method for black hole mass measurement is that of 
spatially resolved spectroscopy of rotating gas disks in the centers of 
nearby galaxies. Because spatial resolutions of typically 0\farcs1 are needed,
this method has mainly utilized the Hubble Space Telescope (HST) 
\citep[e.g.][]{ffj96,mb98,verdoes-kleijn00,barth01,marconi03,atkinson05}.
With the superior resolution of the HST, gas 
rotation close to, or inside, the black hole's sphere of influence, i.e.\
inside which the black hole potential dominates over the bulge potential,
can be measured. For the more massive black holes, say $>10^{8}$ 
M$_{\odot}$, or for sufficiently nearby galaxies, the sphere of influence can 
also be resolved from the ground at near-infrared wavelengths
\citep[e.g.][]{marconi01,tadhunter03}. 

Now that the Space Telescope Imaging Spectrograph (STIS) on board HST is no
longer operational, it becomes necessary to find alternative 
techniques with  
sufficient spatial resolution to probe dynamics of gas and stars 
close to the black hole's sphere of influence. Adaptive optics assisted 
spectroscopy in the near-infrared is the most obvious alternative
because resolutions of $<0\farcs1$ can routinely be reached 
\citep{haring-neumayer06}. Here we show that also diffraction limited 
spectroscopic imaging in the mid-infrared at 8m class telescopes is a viable 
technique.

Some galaxies have nuclei that are hidden behind large amounts of dust, hence
the nuclear regions become unavailable at optical, and sometimes also at
near-infrared, wavelengths. However, the dust is penetrated at mid-infrared 
wavelengths, and mid-infrared spectroscopy at high spatial resolution can 
therefore be used to constrain black hole masses in dust enshrouded galaxy 
nuclei. The starburst/Seyfert~2 galaxy \object{NGC 7582} 
is an example of a such a 
galaxy. Here we present data of this galaxy taken with the ESO mid-infrared 
instrument VISIR (VLT Imager and Spectrometer for mid-InfraRed). The data 
consist of a narrow-band image and two high-resolution long-slit spectra 
centered on the [\ion{Ne}{ii}]12.8$\mu$m line. 

NGC 7582 is a highly inclined barred spiral galaxy (SBab) at a systemic 
velocity $cz\approx 1620$ km\,s$^{-1}$ with a composite starburst/Seyfert~2  
nucleus. The nucleus has a well-defined one-sided ionization cone in
[\ion{O}{iii}]5007 \citep{storchi91}. 
In the center, there is circum-nuclear H$\alpha$ emission 
distributed in a rotating kpc-scale disk in the plane of the galaxy, 
perpendicular to the cone axis \citep{morris85}.
The nucleus is hidden behind large amounts
of dust \citep{regan_mulchaey99}, presumably associated with 
the putative torus. 

The starburst in the center is associated with the nuclear kpc-scale 
disk seen in low-excitation emission lines \citep{morris85,sosabrito01} and in
the soft X-rays, extended emission from the starburst is seen \citep{lwh01}. 
Low-resolution spectra and images in the mid-infrared show PAH (Polycyclic 
Aromatic Hydrocarbon) emission from star forming regions 
\citep{siebenmorgen04}. The nuclear AGN (Active Galactic Nucleus) point source
is strong and well-defined in the $L$ (3.8$\mu$m) and $M$ (4.7$\mu$m) bands, 
and the star forming kpc-scale disk also shows up in the $L$-band image 
\citep{prieto02}, probably because of PAH emission at 3.3$\mu$m. Near-infrared
spectra show both broad and narrow Br$\gamma$ \citep{reunanen03}. \citet{wg06}
discuss the star formation in the disk in more detail, in particular the 
presence of embedded star clusters discovered in the 
[\ion{Ne}{ii}] narrow-band image, with no counterparts detected at optical 
or near-infrared wavelengths. They estimate that the two brightest sources
in the disk may contain roughly 1500 O stars each, and are likely to be 
young, super star clusters embedded in their own dusty birth material. 

Throughout the paper, a Hubble constant of $H_{0}=70$ km\,s$^{-1}$\,Mpc$^{-1}$
is used, hence one arcsec corresponds to $\approx108$ pc in the 
rest-frame of the galaxy.

\section{Observations and data reduction}
\label{section:data}

A narrow-band image and two long-slit spectra were obtained 
with the VISIR instrument mounted at the Cassegrain focus of the UT3 Melipal
telescope. The VISIR detector is a SiAs 256$\times$256 DRS array, and the 
pixel scale used during the observations was 0\farcs127, yielding a field of 
view of 32\farcs5$\times$32\farcs5. 

The image was taken on 2004 September 29 through the [\ion{Ne}{ii}] 
narrow-band filter centered on 12.81$\mu$m (half-band width 0.21$\mu$m). A
perpendicular chopping-nodding technique was used with a chop throw of 
12\farcs0 and a total integration time of one hour. The two high-resolution 
spectra were obtained on
October 2 using the [\ion{Ne}{ii}]12.8$\mu$m long-slit echelle mode. Both 
spectra are centered on the nucleus. One spectrum was obtained with the slit 
approximately parallel to the major axis of the circum-nuclear disk and the 
other with the slit perpendicular to the major axis. A parallel chop-nod 
technique 
was used with nodding along the slit. The chop throw amplitude was 12\farcs0 
and the total integration time for each spectrum was one hour.

The spectra were obtained with the 0\farcs75 slit and the grating tilted to a 
central wavelength of 12.8803$\mu$m. With this setting, the dispersion is 
1.372$\times$10$^{-4}\mu$m per pixel, and the resolving power $R\approx16000$. 
For calibrations, a telluric standard star, HD1522, with spectral type K1.5III,
was observed. Conditions on both nights were clear with humidity below 10\%.

The PSF in the narrow-band image has a FWHM (Full Width at Half Maximum) of 
0\farcs4. The spatial 
resolution in the spectrum is also 0\farcs4, hence the data are diffraction
limited as the diffraction limit for the VLT is 0\farcs39 at 12.8$\mu$m.
The spectral resolution is given by the width of the function which is a 
convolution of the Line Spread Function (LSF) and the slit function. If we 
assume that the slit function is a 
rectangular function with width equal to the slit width of 5.9 pixels, and 
convolve it with a Gaussian LSF with a FWHM of 4 pixels, we obtain a profile
with a width of $\approx 7$ pixels, corresponding to the FWHM of sky lines
in the spectra. The LSF can therefore be 
approximated by a Gaussian with a FWHM of 4 pixels, implying that the 
instrumental broadening (FWHM) is 4 pixels, or $\approx13$ km/s.

Pipeline-reduced images and spectra are used throughout the paper.
For each telescope nodding position the average of 
all chopped frames is built and the different nodding frames are combined 
to derive the final image. The wavelength calibration is performed by 
cross-correlating the position
of skylines in the observed spectrum with that of a HITRAN 
spectrum \citep{hitran}. The pipeline wavelength calibration was checked
by identifying the three strongest skylines with the corresponding skylines 
in high resolution ($R\approx 130,000$) $N$-band solar spectra from the 
McMath Fourier Transform Spectrometer \citep{wallace94}, and an agreement to 
within less than two pixels was found. 

\section{The circum-nuclear star forming disk}
\label{section:nbimg}

\begin{figure}
\resizebox{\hsize}{!}{\includegraphics{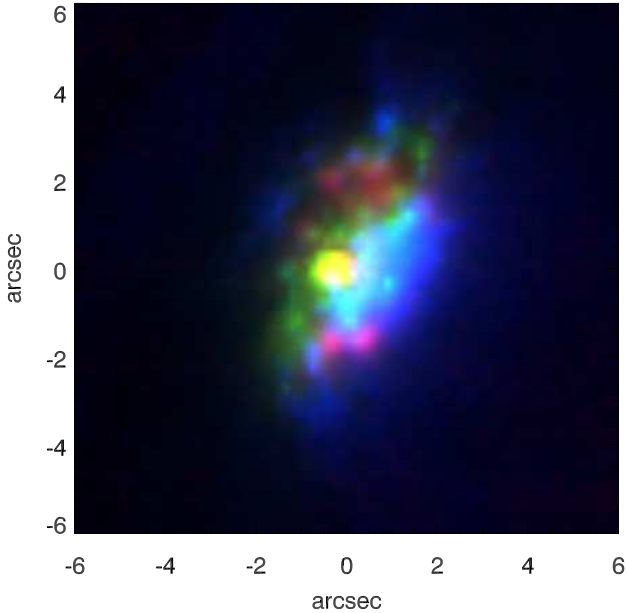}}
\caption{Color composite of the central kpc.
The optical WFPC2 $F606W$ image is coded as blue and the 
near-infrared NICMOS $F160W$ image as green. The narrow-band image
centered on the [\ion{Ne}{ii}]12.8$\mu$m line is coded in red. 
North is up and East is to the left.}
\label{figure:colim}
\end{figure}

In order to study the structure of the circum-nuclear disk, we 
formed a color image of the central $\sim10\arcsec \times10\arcsec$ by 
combining the VISIR narrow-band image with archival HST  WFPC2 and NICMOS 
images taken through the $F606W$ and $F160W$ filters (programmes 8597 and 7330,
respectively). The NICMOS image was obtained with the NIC2 detector, and for 
the WFPC2 image, we used only the PC chip covering the nucleus. The colour 
image is shown in Fig.~\ref{figure:colim} where the WFPC2 image has been 
coded in blue, the NICMOS image ($H$-band) 
in green and the [\ion{Ne}{ii}] narrow-band image in red. The image agrees well
with the model favoured by \citet{morris85} based on observations of 
[\ion{O}{iii}] and H$\alpha$. They suggest a rotating, star forming kpc-scale 
disk surrounding the nucleus with a one-sided outflowing ionization cone with 
the axis perpendicular to the disk plane. 
 
There is a striking anti-correlation between the three colors in the image.
The blue-coded optical emission dominated by the stellar continuum is obscured 
by a dust band stretching across the nucleus from the South-East 
towards the North-West. Since there is some contribution from H$\alpha$ to the 
optical flux the cone-structure can be seen in blue at 
$PA\approx250\degr$ opening up towards the West.
The green-coded near-infrared emission coming mainly from stars 
in the circum-nuclear disk is seen through the dust where the 
optical emission is obscured. But even the near-infrared emission is 
patchy because of dust obscuration, and in regions where both the near-infrared
and optical emission has been extinguished by dust, 
the red-coded mid-infrared emission peeks through. 

[\ion{Ne}{ii}] is collisionally excited and the 12.8$\mu$m emission
traces gas that has been ionized 
by UV radiation from newly formed, massive stars. 
Several knots of 
emission are seen, associated with dense, embedded regions of star formation. 
\citet{wg06} discuss the possibility that the bright [\ion{Ne}{ii}] knots
may be young star clusters consisting of $\sim 10^{3}$ stars, 
embedded in dusty material.
The two most prominent knots are seen to the South of the nucleus.
The rightmost knot is seen in the deconvolved 11.9$\mu$m image (tracing 
PAH emission) by \citet{siebenmorgen04}, but the one to the left 
is undetected at 11.9$\mu$m. It is also very weak, or undetected, in H$_{2}$ 
\citep{sosabrito01}. 

Using isophotal ellipse fitting, we measure the ellipticity\footnote{Here 
defined as $e \equiv 1 - \frac{b}{a}$ with $a$ and $b$ being major and minor 
axis, respectively} and position angle of the [Ne{\sc ii}] disk after masking 
out the knots. The best measurements at a semi-major axis of 
$\approx$ 2\farcs5 give an ellipticity of $0.47\pm0.11$ and a position angle 
of $158\degr\pm4\degr$. Assuming a thin, circular disk, the inclination angle 
is given by $\cos i = \frac{b}{a}$, implying that $i=58\degr$. The 
circum-nuclear gas disk therefore has, within the errors, the same inclination
angle as the galaxy, $60\degr\pm10\degr$ \citep{morris85}, and is
consistent with lying in the plane of the galaxy. The rotation of the disk 
follows the direction of rotation of the galaxy. 


\section{The velocity field in the disk}

\begin{figure}
\centering
\includegraphics{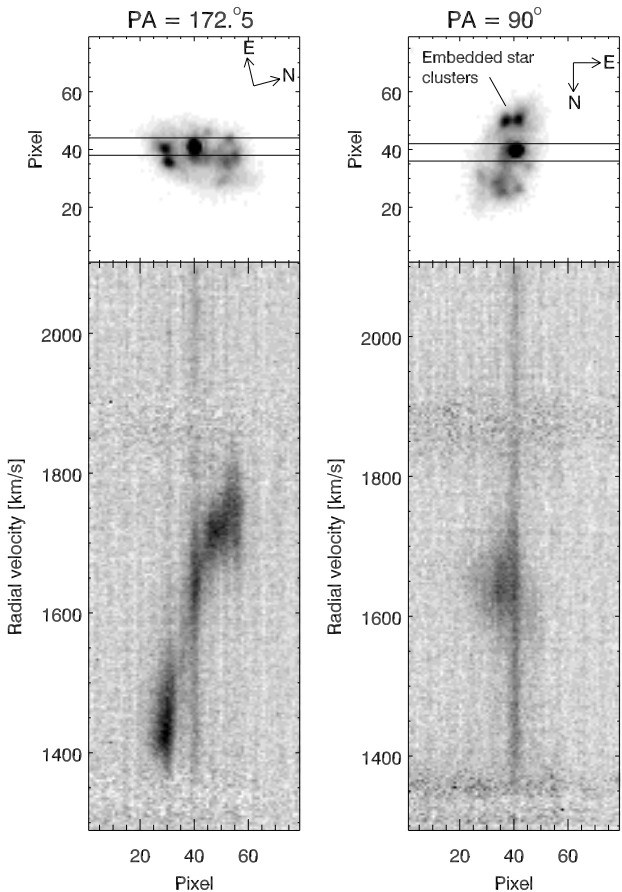}
\caption{The two high-resolution spectra of the [\ion{Ne}{ii}] line with the 
slit positions indicated on the narrow-band image. The parallel spectrum is 
shown to the left, and the perpendicular spectrum to the right. The two 
regions of increased noise at $\approx1350$ and 1850 km\,s$^{-1}$ are due to 
strong skylines. Continuum from hot dust emission in the center is seen.}
\label{figure:im+spec}
\end{figure}

The two [\ion{Ne}{ii}] longslit spectra show the velocity field across the disk
along two directions almost perpendicular to each other. 
Fig.~\ref{figure:im+spec} shows the full two-dimensional spectra together
with the narrow-band image with slit positions indicated. The parallel spectrum
(shown to the left) 
was taken with the slit positioned over the nucleus and one 
of the compact knots of [\ion{Ne}{ii}] emission to the South. The position 
angle is $PA=172\fdg5$, which is close to the position angle of the major axis 
of the disk. The perpendicular 
spectrum, shown in the right-hand panel, was obtained at 
$PA=90\degr$, i.e.\ along the axis of the ionization cone. Both spectra 
show continuum emission from hot dust in the centre. The parallel 
spectrum clearly exhibits rotation, whereas the velocity across the minor axis 
as seen in the perpendicular spectrum is roughly constant. 

\begin{figure}
\centering
\includegraphics{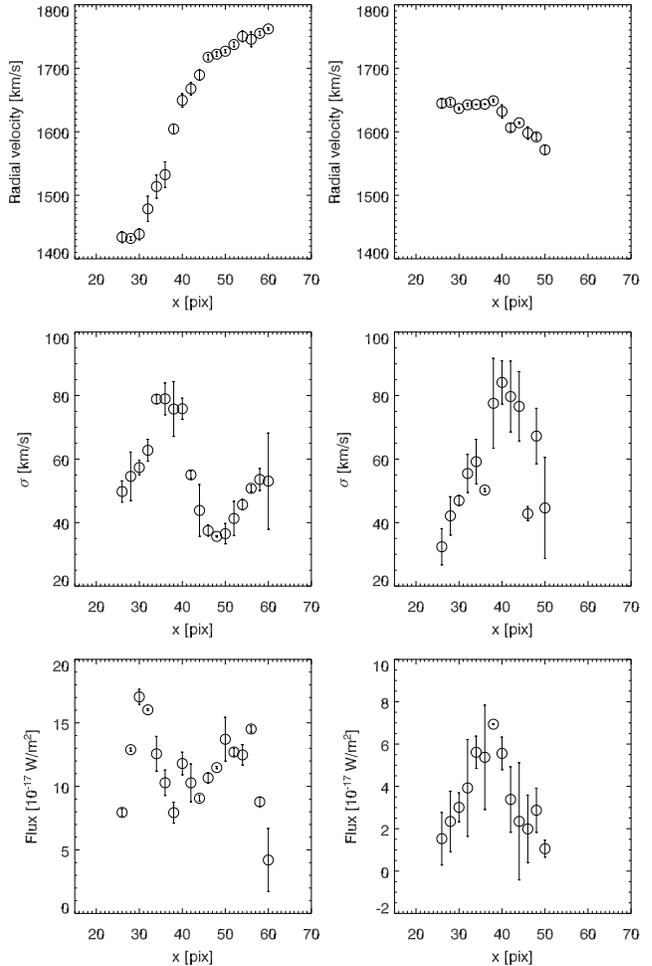}
\caption{Radial velocity, velocity dispersion and flux of 
the [\ion{Ne}{ii}] line as a function of position along the slit in the 
two spectra. The panels to the left show the parallel spectrum,
and the panels to the right, the perpendicular spectrum.}
\label{fig:velfield}
\end{figure}

In order to derive velocity curves, apertures two pixels wide were 
extracted from the spectra. A Gaussian was fitted to each extracted spectrum 
so that the line centroid, FWHM and flux as a function of position along the 
slit could be determined. The results are shown in Fig.~\ref{fig:velfield} 
where wavelength has been converted to radial velocity and FWHM to velocity
dispersion (FWHM/2.35). 

The parallel chop-nod technique produces one positive and two 
negative images on the detector. The positive image contains half of the total
flux and each of the two negative images, a quarter of the total flux. 
As we were concerned with determining the position of the wavelength
centroid as accurately as possible we chose not to combine the three spectral
images, as imperfect registration could have introduced centroid shifts. 
Averaging the three images also does not increase the signal-to-noise by more 
than a factor of $2/\sqrt{3}$ (because each image contains only part of the 
total flux). 

The data shown in 
Fig.~\ref{fig:velfield} were extracted from the positive spectrum which 
has the highest signal-to-noise. 
For estimating uncertainties in wavelength centroid, FWHM and line
flux, we utilized both the positive and the two negative spectra. The 
uncertainty in each quantity $x$, 
where $x$ may be line centroid, FWHM or flux, was calculated
as $(|x_{\rm pos} - \bar{x}_{\rm neg}|)/2$, where $\bar{x}_{\rm neg}$ is 
the average of the measurements from the two negative images and 
$x_{\rm pos}$ is measured from the positive image. We find an uncertainty 
in the wavelength centroid of typically 5--10 
km\,s$^{-1}$, depending somewhat on the position along the slit. There is an 
additional uncertainty due to the fact that the spatial resolution in the
spectrum is smaller than the slit width. 
However, this is a systematic offset which does not vary with wavelength, 
hence was not taken into account here.
The uncertainty in the velocity dispersion is 3--7 km\,s$^{-1}$ and the 
uncertainty in the flux, from a few up to $\approx10$ per cent. 

The rotation curve in the upper left-hand panel of Fig.~\ref{fig:velfield} 
shows that the disk rotates with an amplitude of $\approx 200$ km\,s$^{-1}$.
The rotation curve also appears relatively smooth with
some irregularities close to the edges. In the blueshifted southern part of 
the disk (over pixels 26--38) a slight turn-over in the velocity is observed, 
probably indicating that most of the mass probed by the rotation curve is 
contained in the probed region. The rotation of the disk is such that the 
emission to the North is redshifted and the emission to the South blueshifted. 

The perpendicular velocity curve to the right shows a more constant velocity 
of $\approx1650$ km\,s$^{-1}$ across the disk with a slight decrease in 
velocity toward the East. An approximately constant velocity profile is 
what we expect from a perpendicular spectrum aligned along the minor axis of 
a thin rotating disk. 

The middle row of panels in Fig.~\ref{fig:velfield} shows
velocity dispersion as a function of position along the slit. Typically 
we measure dispersions of 40--80 km\,s$^{-1}$ with maximum around the 
center of the disk. There are several different factors contributing 
to the velocity dispersion. Apart from unresolved rotation, there are 
contributions from instrumental and thermal broadening. 
The instrumental broadening is $\approx 13$ km\,s$^{-1}$ (see 
Section~\ref{section:data}), and the thermal broadening broadening (FWHM) for
a $10^{4}$ K disk $\approx20$--30 km\,s$^{-1}$. This adds up to 
a FWHM of $\approx24$--33 km\,s$^{-1}$, or 
$\sigma\approx10$--14 km\,s$^{-1}$. We measure $\sigma > 40$ 
km\,s$^{-1}$ so there must be a non-negligible contribution from bulk motion 
in the gas. The increase in velocity dispersion toward the center of 
the disk is seen in both the parallel and the perpendicular spectrum, 
probably caused by spatially unresolved rapidly rotating gas close to the 
nucleus.

The radius of the black hole sphere of influence is 
$GM_{\rm BH}/\sigma_{*}^{2}$ where $M_{\rm BH}$ is the black hole mass and 
$\sigma_{*}$ is the central stellar velocity dispersion. The velocity 
dispersion of NGC 7582 listed in the HyperLeda 
database\footnote{http://leda.univ-lyon1.fr/} is 
157$\pm$20 km/s, and using aperture 
corrections by \citet{jorgensen95} and the $M_{\rm BH}-\sigma_{*}$ relation by 
\citet{tremaine02}, we estimate a black hole mass of 
$\approx 4\times10^{7} M_{\sun}$. Hence the radius of the black hole sphere of 
influence is $\la $ 0\farcs1. The spatial resolution in the [\ion{Ne}{ii}] 
spectrum is a factor of 3--4 larger than this, so the sphere of influence
is unresolved. 
But even though the spatial resolution is not sufficient to probe
gas rotation inside the sphere of influence, an estimate of the black 
hole mass can be obtained from the steepness of the rotation curve across the 
nucleus. In the following, we set up a model of the rotating disk to be able 
to estimate the central black hole mass. Our modeling follows largely that 
of \citet{barth01}, and is the standard gas-dynamical method used for nuclear 
gas disks in galaxies 
\citep[see also][]{marconi01,tadhunter03,atkinson05}. The gas 
disk is assumed to rotate in the combined potential of the galaxy bulge and 
the central supermassive black hole. By varying the black hole mass as well as
other parameters such as disk inclination and mass-to-light ratio, different 
models can be compared to the data in order to find the best fit. We describe 
the modeling in more detail in the following section.

\section{The model velocity field}

\subsection{The basic equations}
\label{section:basic}

We assume that the circum-nuclear disk can be approximated by a flat, circular
disk with constant inclination rotating in the gravitational potential $\Psi$. 
At every radius $R$ in the disk, the rotation is dictated by 
a combination of the gravitational potential from the stellar bulge and
the central, supermassive black hole,
\begin{equation}
\Psi(R) = \Psi_{*}(R) + \frac{GM_{\rm BH}}{R}.
\label{equation:vc}
\end{equation}
\noindent
The bulge potential is $\Psi_{*}$ and the black hole point mass potential
is $GM_{\rm BH}/R$. The bulge potential dominates far 
out in the disk, but close to the center, within the black hole sphere of 
influence, there is a non-negligible contribution from the black hole. 
The circular velocity, $v_{\rm c}$, in the disk is obtained by differentiating
the potential with respect to $R$:
\begin{equation}
v_{\rm c}^{2}(R) = R \frac {\partial{\Psi}} {\partial{R}}. 
\label{equation:vcirc}
\end{equation}

The plane of the sky is defined by the coordinate system 
$(x^{\prime},y^{\prime})$ where the $x^{\prime}$-axis coincides with the major
axis of the disk. We also define the slit coordinate system 
$(x_{\rm s},y_{\rm s})$ with the $y_{\rm s}$-axis 
along the length of the slit.  The transformation between the two coordinate 
systems is given by 
\begin{eqnarray} 
x^{\prime} & = & (x_{\rm s}-b)\sin\theta  + y_{\rm s}\cos\theta \nonumber \\
y^{\prime} & = & -(x_{\rm s}-b)\cos\theta + y_{\rm s}\sin\theta 
\end{eqnarray}
\noindent
\citep[see appendix~B by][]{marconi01}, where $\theta$ is the angle between 
the slit and the major axis of the disk. In our case, the slit is rotated 
by $\theta = -14\fdg1$ with respect to the major axis
(i.e.\ 158\fdg4, which is the position angle of the disk, minus the slit 
position angle of 172\fdg5). 
The center of the disk, i.e.\ the dynamical center, is assumed to have 
coordinates $x_{\rm s}=b$ and 
$y_{\rm s}=0$, hence the $b$-parameter enables us to take into account that the
dynamical center may not be at the exact center of the slit. 

The inclination of the disk is $i$, and we have by deprojection that 
\begin{equation}
R^{2} = x^{\prime 2} + \left(\frac{y^{\prime}}{\cos i}\right)^{2}.
\end{equation}
\noindent
The velocity along the line of sight is given by 
\begin{equation}
v = v_{\rm c} \sin i \frac{x^{\prime}}{R}.
\end{equation}
From the equations above, we calculate the velocity along the line of sight
in the slit coordinate system for any combination of bulge potential, black 
hole mass, disk inclination and dynamical center position $b$. By comparing the
velocity curves derived from the different models to the observed one, the 
best-fitting parameters, including black hole mass, can be found.

\subsection{The bulge gravitational potential}
\label{section:bulgepot}

An important part of setting up the model is to determine the bulge 
gravitational potential, $\Psi_{*}(R)$. In order to do this, we use the stellar
density in the bulge as a tracer of the gravitational potential. The NICMOS
$F160W$ image is well suited for this because the near-infrared is a good 
tracer of stellar mass and because the image has high spatial resolution. 
Moreover, near-infrared is less affected by dust than the 
optical, a clear advantage here because the center of NGC~7582 is 
strongly affected by dust. 

In order to parameterize the surface brightness distribution, we use 
the Multi-Gaussian Expansion (MGE) method developed by \citet{cappellari02}. 
It consists of parameterizing the galaxy surface brightness with a 
series expansion of two-dimensional Gaussian functions which can be converted 
to a gravitational potential $\Psi_{*}$.
But since even the $F160W$-band is affected by dust, we found 
it necessary to make a correction before applying the MGE analysis to the 
image. This was done by mapping the extinction across the central parts of the 
galaxy by forming an $F606W-F160W$ image. The absolute extinction in 
the $F160W$-band was thereafter calculated by assuming a galactic reddening 
law \citep{cardelli89} and a correctly-normalized dust-corrected image made by 
matching pixels in the uncorrected image to pixels in the corrected image that 
appeared unreddened. For simplicity, we assume a uniform intrinsic colour in 
the bulge, but in reality a gradient may exist.

In order to avoid introducing too many distortions in the surface brightness 
profile, the dust lane that runs across the galaxy was masked out before
performing the MGE fit. The parameters of the best fit 
is listed in Table~\ref{table:MGEfits}, where each Gaussian is characterized 
by a surface density $I_{k}^{\prime}$, a dispersion $\sigma_{k}^{\prime}$, and 
an axial ratio $q_{k}^{\prime}$. The 
luminosity $L_{k}$ in the last column is the total luminosity of the 
Gaussian. Primed quantities are projected whereas unprimed quantities are 
intrinsic. The luminosity $L_{k}$ is calculated as 
$2 \pi I^{\prime}_{k} q_{k}^{\prime} \sigma_{k}$, where we have assumed that
the gravitational potential is axissymmetric so that 
$\sigma_{k}^{\prime} = \sigma_{k}$.

In order to take the shape of the NICMOS PSF into account in the MGE 
analysis, we also computed a TinyTim PSF \citep{kh97} and parameterized
it with four circular Gaussians \citep{cappellari02}:
\begin{equation}
{\rm PSF}(R^{\prime}) = \sum_{k=1}^{M} G_{k}\exp[-R^{\prime 2}/(2\sigma_{k}^{*2})]/(2\pi \sigma_{k}^{*2}),
\end{equation}
where $\sigma^{*}_{k}$ are the dispersions and $G_{k}$ the 
weights of the Gaussians. The parameters of the MGE 
fit to the NICMOS PSF are listed in Table~\ref{table:tinytim}. 

The gravitational potential corresponding to an MGE density profile in the 
oblate axissymmetric case is given by 
\begin{equation}
\Psi_{*}(R,z=0) = G\Upsilon\sqrt{\frac{2}{\pi}} \sum_{k=0}^{N} 
  \frac{L_{k}\ \begin{cal}Q\end{cal}_{k}(R,z=0)}{\sigma_{k}}
\label{eq:mgepotential}
\end{equation}
\noindent
\citep{emsellem94,cappellari02a},
where $\Upsilon$ is the stellar mass-to-light ratio. 
The radius in the equatorial plane is $R$ and $z=0$ because the circular 
velocity in the equatorial plane is determined solely by the radial component 
of the force. We assume that the gas disk lies in the equatorial plane of 
the galaxy, hence the radial velocities we derive also lie in this 
plane.
The factor $\begin{cal}Q\end{cal}_{k}$ is defined as 
\begin{equation}
\begin{cal}Q\end{cal}_{k}(R,z=0) = \int_{0}^{1} 
  \frac{\exp(-R^2 T^2/2\sigma_{k}^2)} {\sqrt{1-(1-q_{k}^2)T^{2}}}{\rm d}T, 
\end{equation}
\noindent
(see Cappellari 2002) 
where $q_{k}$ is the intrinsic axial ratio,  calculated under the 
assumption that the galaxy has an inclination of 56\fdg2, as measured by the 
MGE fitting routine.

\begin{table}
\caption{Parameters of the MGE fit to the galaxy bulge in the NICMOS
F160W filter.}
\begin{center}
\begin{tabular}{lllll}
\hline
$k$ & $I^{\prime}_{k}$ & $\sigma^{\prime}_{k}$  & $q^{\prime}_{k}$ & $L_{k}$ \\
    & L$_{\odot,F160W}$\,pc$^{-2}$ & arcsec & & $10^{6}$ L$_{\odot,F160W}$ \\
\hline
1 & 33307 & 0.330 & 0.716 &  192.5 \\
2 & 1041  & 1.333 & 0.872 &  119.5 \\
3 & 4969  & 2.229 & 0.556 & 1018.5 \\
4 & 343   & 7.188 & 0.556 &  730.2 \\
\hline
\end{tabular}
\end{center}
\label{table:MGEfits}
\end{table}

\begin{table}
\caption{Parameters of the MGE fit to the NIC2 PSF in the F160W filter.}
\begin{center}
\begin{tabular}{lll}
\hline
$k$ & $G_{k}$ & $\sigma_{k}^{*}$ [pix] \\
\hline
1 & 0.251 & 0.75 \\
2 & 0.475 & 1.78 \\
3 & 0.146 & 7.64 \\
4 & 0.127 & 19.51 \\
\hline
\end{tabular}
\end{center}
\label{table:tinytim}
\end{table}

\subsection{Generation of two-dimensional synthetic spectra}

Having determined the bulge gravitational potential, we use the equations 
outlined in Section~\ref{section:basic} to calculate models for different 
values of black hole mass and dynamical center offset, $b$. The disk 
inclination and the mass-to-light ratio are held constant at
$i=58$\degr\ and $\Upsilon = 3.8$ as the data do 
not have sufficient resolution to allow for a multi-parameter fit. 
The mass-to-light ratio is fit separately, prior to fitting $M_{\rm BH}$
and $b$, as explained in Section~\ref{section:bhmass_constraints}.

Each model is ``observed'' with the 0\farcs75
slit and the instrumental setup that was used during the observations.
We also convolve with the resolution of the observations, and fold the 
[\ion{Ne}{ii}] narrow-band image into the model in order to produce a 
two-dimensional synthetic spectrum. 
We follow the procedure of \citet{barth01}, which is briefly described below. 
A detailed explanation can be found in Barth et al.'s paper. 

First we set up a coordinate grid on which to compute the synthetic spectrum. 
The grid is subsampled by a factor of two compared to the data, so the 
pixel size in the model spectrum is half the size of the real pixels. The 
model is therefore resampled before comparing to the data.  
The construction of the synthetic spectrum consists of calculating the observed
surface brightness of the disk through the 0\farcs75 slit. This is done by
assuming that the intrinsic line profiles are 
Gaussians and adding them up at every point in the disk, weighing by the 
[\ion{Ne}{ii}] surface brightness (as provided by the narrow-band image) 
and smearing by the seeing PSF.  

The synthetic line profile, $f_{y_{\rm s}}(v)$, is given by Barth et al.'s 
(2001) eq.~4,
and is reproduced here in order to clarify the discussion:
\begin{eqnarray}
f_{y_{\rm s}}\left(v\right) & = & \sum_{x_{\rm s}=x_{0}}^{x_{1}} \sum_{i,j} S_{ij} P(i,j|x_{\rm s},y_{\rm s}) \nonumber \\
         & & \times \exp \left[\frac{\left[v-v_{\rm p}(i,j)-(x_{\rm s}-x_{\rm c})v_{\rm d}\right]^{2}}{-2\sigma_{\rm r}^{2}}\right].
\label{equation:2dspec}
\end{eqnarray}
The [\ion{Ne}{ii}] surface brightness in the disk at pixel $(i,j)$ on the
model grid is denoted $S_{ij}$, and the value at 
$(x_{\rm s}, y_{\rm s})$ of a PSF centered at pixel $(i,j)$ is denoted 
$P(i,j|x_{\rm s},y_{\rm s})$. Hence the multiplication of $S$ and $P$ 
represents a convolution of the surface brightness distribution with the PSF 
and takes into account that different parts of the disk contribute to 
the velocity profile at a given position. We assume that the PSF is a Gaussian
with FWHM equal to the seeing. During the computations the PSF was truncated 
at $2.5\times$FWHM. Since we are effectively convolving the narrow-band image
with the PSF, the narrow-band image first had to be deconvolved. This was 
achieved by block replicating the original image by a factor of two and 
deconvolving using the Multiwavelength Deconvolution Method
\citep{sm02}. 

Since the intrinsic line profiles are assumed to be Gaussians, the exponential
factor in Eq.~\ref{equation:2dspec} 
contains the model velocity field $v_{\rm p}(i,j)$ and a correction 
factor $(x_{\rm s}-x_{\rm c})v_{\rm d}$ taking into account the 
uncertainty in velocity caused by the fact that the recorded wavelength 
of a photon depends on where it enters the slit along the $x_{\rm s}$-axis. 
The pixel size in velocity units is 
$v_{\rm d} = 3.2$ km\,s$^{-1}$, and the center of the slit is denoted $x_{c}$.
The velocity dispersion of the gas, $\sigma_{\rm r}$, occurs in the 
denominator. As seen from Fig.~\ref{fig:velfield} 
it varies significantly across the disk, hence is difficult to fit by a 
model. We therefore approximate $\sigma_{\rm r}$ with $\approx55$ 
km\,s$^{-1}$ which is the average across the disk. 

In order to avoid including continuum emission in the synthetic spectra, the 
narrow-band image was continuum subtracted before incorporating it into the 
model. Because it was obtained solely for the purpose 
of aiding with the determination of slit positions, images in the continuum 
filters bracketing the [\ion{Ne}{ii}] line were not taken. We therefore 
subtract a continuum based on that measured in the spectrum instead, 
achieved by first scaling a point source to the continuum flux measured in 
the spectrum, and thereafter subtracting it from the image.
Because the narrow-band image is folded into the model, the position of the 
slit relative to the narrow-band 
image had to be determined. This was done by first rotating the image so 
that the length of the slit was parallel to the y-axis of the rotated
image, and thereafter collapsing the spectrum in the wavelength direction and 
matching it to the section of the image covered by the slit. 

\begin{figure}
\begin{center}
\resizebox{\hsize}{!}{\includegraphics{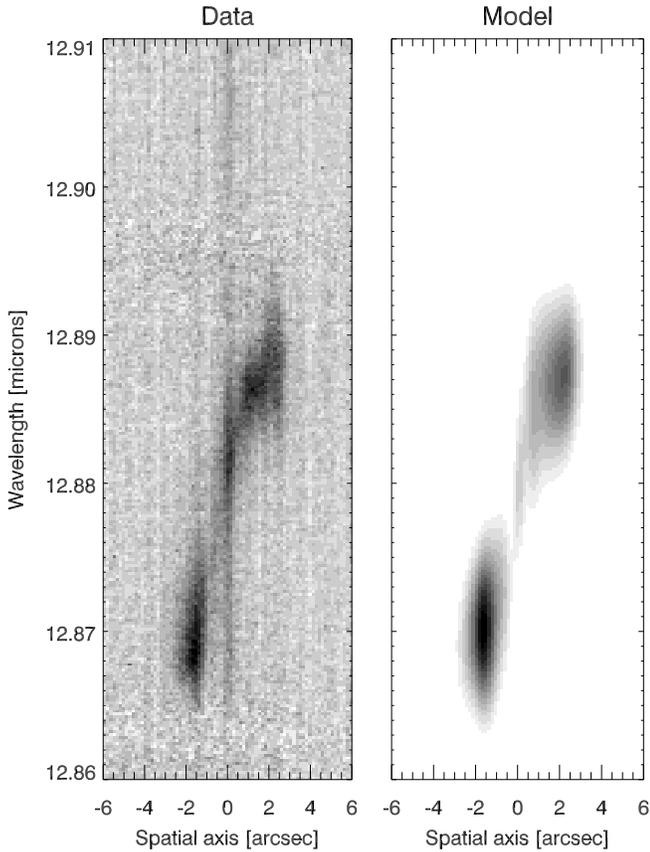}}
\caption{Comparison between the real spectrum (left) 
and a synthetic spectrum (right), calculated for a black hole mass of 
$5.5\times10^{7}$ M$_{\odot}$ and with the dynamic center shifted 0\farcs3 
from the center of the slit.} 
\label{fig:modelfig}
\end{center}
\end{figure}

The summation over $x_{\rm s}$ in Eq.~\ref{equation:2dspec} runs across the 
width of the slit, where the 
edges of the slit are given by $x_{\rm s} = x_{0}$ and $x_{\rm s} = x_{1}$.
The $y_{\rm s}$-coordinate runs along the spatial direction of the spectrum. 
To create the full two-dimensional spectrum, $f_{y_{\rm s}}(v)$ 
has to be evaluated for every pixel along the spatial direction. 
An example of a synthetic spectrum is shown in Fig.~\ref{fig:modelfig}. The 
two regions of emission on either side of the center are well 
reproduced, as are their relative compactness and brightness. As the 
synthetic spectrum is noise free, and also void of sky lines, it differs 
slightly from the observed one, especially around $12.863\mu$m where there is 
increased noise due to the subtraction of a sky line. 
Also, since the synthetic spectrum was produced by convolving with a 
continuum-subtracted narrow-band image, the two differ in flux at the center. 

In order to derive model rotation curves, we made extractions from the 
synthetic two-dimensional spectra in a manner identical to that which was done
for the real spectrum. 

\subsection{Constraints on the black hole mass}
\label{section:bhmass_constraints}

Given that the VISIR spectrum has a spatial resolution of 
$\approx 0\farcs4$, we merely fit two 
parameters, namely the black hole mass and the location of the dynamical
center within the slit. The inclination of the disk is fixed
at $i=58\degr$ and the mass-to-light ratio at $\Upsilon=3.80$.
We calculate models on a grid of black hole mass and dynamical center
location, $b$.
The spacing in $M_{\rm BH}$ was varied between 0.5$\times$10$^{7}$ and 
1$\times$10$^{7}$ M$_{\odot}$, and the spacing in $b$, between 0\farcs1 and 
0\farcs05, depending on how close the model was to the 
global minimum in $\chi^{2}$.

\begin{figure}
\resizebox{\hsize}{!}{\includegraphics{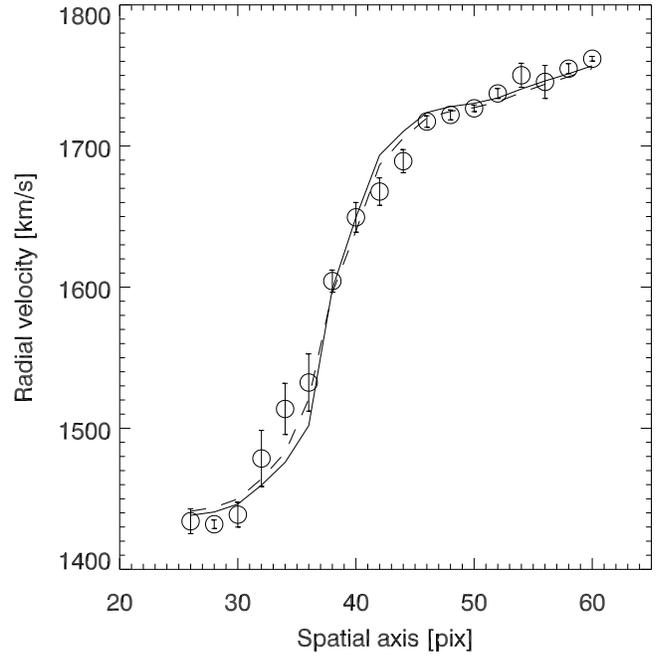}}
\caption{The open circles show the observed rotation curve and the solid line 
indicated the best-fit model with $M_{\rm BH} = 5.5 \times 10^{7}$ 
M$_{\odot}$ and $b=$0\farcs3. The dashed line corresponds to a model with 
$b=0.3$ and $M_{\rm BH}=0$.}
\label{fig:vcfig}
\end{figure}

For every pair of $M_{\rm BH}$ and $b$ a synthetic two-dimensional spectrum
was constructed, and one-dimensional spectra extracted from it using apertures
of the same width as for the real data. Thereafter a Gaussian fit was made to 
each in order to determine the line centroid so that a radial velocity 
curve could be derived. The best-fitting
model was found in a standard manner by minimizing $\chi^{2}$ between the 
observed and the model rotation curve. Fig.~\ref{fig:vcfig} shows the
observed rotation curve together with a model rotation curve extracted from 
a synthetic spectrum.

In order to convert the MGE surface density profiles to a gravitational
potential, the stellar mass-to-light ratio, $\Upsilon$, in the bulge has to 
be known, see Eq.~\ref{eq:mgepotential}. Outside the black hole's sphere of 
influence, the rotation curve does not depend on either the black hole mass 
or the exact location of the dynamical center in the slit. Hence
the outer parts of the rotation curve are not sensitive to $M_{\rm BH}$ and
$b$, but depend instead on the mass-to-light ratio.
We therefore fit $\Upsilon$ separately, before fitting $M_{\rm BH}$ and 
$b$, by neglecting the inner
few points on the rotation curve and fitting only the outer parts. Synthetic
spectra were generated with $M_{\rm BH}=0$ and $b=0$ and with mass-to-light 
ratios in the range 2.4--4.8 in steps of 0.2, while keeping the disk 
inclination constant at $i=58$\degr. Apertures were extracted and model 
rotation curves produced. The best-fit mass-to-light ratio was thereafter 
found by fitting the model rotation curves to the observed one with the three 
inner apertures (centered at pixels 36, 38 and 40) excluded. The best-fit 
model has $\Upsilon = 3.80$ M$_{\odot}$/L$_{\odot}$ in the $F160W$ filter, in 
accordance with mass-to-light ratios observed in early-type spiral galaxies 
\citep{moriondo98}. We fix the mass-to-light ratio to 3.80 for the
remainder of the fitting.

Upon comparing the model rotation curve to the observed one we have to decide
on which aperture on the two curves should have the same radial velocity. 
Ideally, this corresponds to the point which has zero velocity in the
rest frame of the galaxy, i.e.\ the point having radial velocity 
equal to the systemic velocity. However, because of the high 
spectral resolution of the VISIR data, the systemic velocity of NGC~7582 known
from the literature is not accurate enough. Instead, we use the perpendicular 
spectrum to help constrain the radial velocity of the dynamical center. 
If the slit for the perpendicular spectrum is aligned along the minor axis of 
the disk, we expect to see a constant velocity along the slit 
corresponding to the systemic velocity. Examining the upper ß
right-hand plot in Fig.~\ref{fig:velfield}, we see that there is a region over 
pixels 26--38 with constant velocity $\approx$1644 km\,s$^{-1}$. If this 
corresponds to the systemic velocity of the galaxy, then the aperture centered 
on pixel 40 in the {\em parallel} spectrum is the correct center to choose 
for the fitting because this is the aperture closest to 1644 km\,s$^{-1}$. 
However, if the region with velocity $\approx1644$ km\,s$^{-1}$ in the
perpendicular spectrum is affected by an outflow or some bulk motion, the center 
could have a lower radial velocity. The aperture centered on pixel 40 in the
perpendicular spectrum has a velocity of 1632 km\,s$^{-1}$, and could therefore
be closer to the dynamical center. Comparing 
again with the parallel spectrum, a velocity of 1632 km\,s$^{-1}$ lies between 
the apertures centered on pixels 38 and 40. So it is likely that the dynamical 
center in the parallel spectrum either coincides with one, or lies somewhere 
between these two apertures. 
The most probable systemic velocity of the galaxy therefore seems to be 
1630--1650 km\,s$^{-1}$. 
Correcting to heliocentric velocity (the correction factor at the time
of observation is $-$16 km\,s$^{-1}$), we therefore find that the 
systemic velocity of the galaxy is 1614--1634 km\,s$^{-1}$, approximately 60 
km\,s$^{-1}$ higher than that reported in the Leda database based on 
optical emission lines. 
Outflows and bulk motion of the gas make it difficult to 
pin-point accurately the aperture corresponding to the systemic velocity. 
Integral field spectroscopy in the near-infrared
might help in this respect, although even in this case, there may be problems 
related to dust obscuration and outflows. 

By requiring that the observed and the modeled rotation curve have the
same velocity in the aperture centered on pixel 38, we derive
a best-fitting model with $M_{\rm BH} = 1.0 \times 10^{8}$ M$_{\odot}$ and 
$b=$0\farcs35 with a reduced $\chi^{2}$ of 3.76. 
By shifting the common aperture to the one centered on pixel 40, the
fit is marginally better with a reduced $\chi^{2}$ of 3.36, 
$M_{\rm BH} = 5.5\times10^{7}$ M$_{\odot}$ and $b=$0\farcs3. 
The slightly different best-fit parameters in each case serves to illustrate 
the level of uncertainty related to the difficulties with knowing the exact 
systemic velocity. The reason we obtain a reduced $\chi^{2}$ greater than 
unity is most likely due 
to outflows and bulk motions in the gas that are not included in the model. 
The current modeling is probably the best
that can be done with these data, and the limitations are due to 
uncertainties related to the spatial resolution being larger than the
sphere of influence of the black hole.

\begin{figure}
\resizebox{\hsize}{!}{\includegraphics{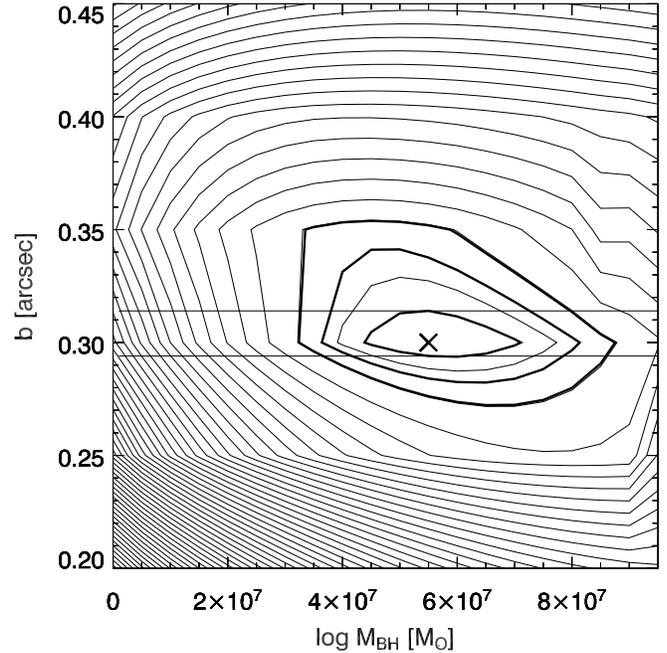}}
\caption{Contours of $\chi^{2}$. The minimum is marked by an 'x',
and the solid lines mark the 68, 95 and 99\% confidence regions.}
\label{fig:chifig}
\end{figure}

We choose the model with $M_{\rm BH}=5.5\times10^{7}$ 
M$_{\odot}$ as our best fit since this has the lowest $\chi^{2}$. This model 
is plotted together with 
the observed rotation curve in Fig.~\ref{fig:vcfig}, and a contour plot of 
$\chi^{2}$ is shown in Fig.~\ref{fig:chifig}, with 68, 95\% and 
99\% confidence regions indicated. The confidence regions on $M_{\rm BH}$ and
$b$ are also listed in Table~\ref{table:chi2}. 

\begin{table}
\caption{The confidence intervals for the parameters
of the best-fitting model having 
$M_{\rm BH} = 5.5\times10^{7}$ M$_{\odot}$ and $b=$0\farcs3.}
\begin{center}
\begin{tabular}{lll}
\hline
        & $M_{\rm BH}$ [10$^{7}$ M$_{\odot}$] & $b$ [arcsec] \\
\hline
68\% CI & [3.2,8.7] & [0.282,0.341] \\ 
95\% CI & [3.6,8.1] & [0.273,0.335] \\ 
99\% CI & [4.4,7.1] & [0.294,0.314] \\ 
\hline
\end{tabular}
\label{table:chi2}
\end{center}
\end{table}

\subsection{The dynamical center}

\begin{figure}
\resizebox{\hsize}{!}{\includegraphics{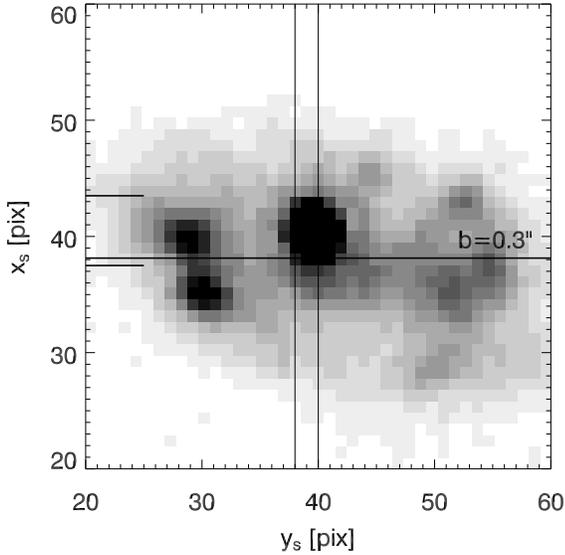}}
\caption{The two shorter horizontal lines to the left
indicate the slit width, and the
two vertical lines the center of the two apertures thought to correspond 
to the systemic velocity of the galaxy. The horizontal line across the image 
indicates the best fit dynamical center position $b=0\farcs3$. The image has 
been rotated, so East is up and North is to the right. 
The axes are labeled according to the slit coordinate system.}
\label{fig:kincenter}
\end{figure}

The best-fitting value $b=0\farcs3$ indicates that the dynamical center may 
not have been exactly centered on the slit when the spectrum was taken, but 
instead located 0\farcs085 from the edge of the slit, west of the center.
We found above that the systemic velocity of the
galaxy probably corresponds to the velocity probed by the apertures centered 
on pixel 38/40 in the parallel spectrum. These constraints can be used to 
find the location of the dynamical center on the narrow-band image, as shown 
in Fig.~\ref{fig:kincenter}. In this figure, the image has been rotated so 
that the length of 
the slit runs horizontally, hence the abscissa is labeled $y_{\rm s}$. The 
ordinate is labeled $x_{\rm s}$ since it runs along the width of the slit. 
The location of the slit is indicated by the two short
horizontal lines on the left-hand side of the image, and the line corresponding
to $b=$0\farcs3 is the horizontal line plotted across the full width of the 
image. The two vertical lines correspond to the center of the two apertures
that were discussed in the previous section as likely to correspond to the 
systemic velocity of the galaxy (between aperture 38 or 40). 
The dynamical center of the galaxy should therefore 
lie in the region where the horizontal line intersects the two 
vertical lines. As seen in Fig.~\ref{fig:kincenter}, this region does not 
correspond to the exact location of the peak in the surface brightness, 
thereby suggesting that the dynamical center may be obscured. 

According to unified models 
\citep{Antonucci93}, the location of the dynamical center is expected to 
coincide with the vertex of the ionization cone because the cone morphology 
is a result of the illumination of gas by the AGN continuum, 
which can only escape within $\pm \sim 45^{\circ}$ of the axis of the torus.
The blue outline of the ionization cone is seen in the colour image
in Fig.~\ref{figure:colim}, but it is difficult to tell from the
image precisely where the vertex is. The location of the dynamical
center might therefore be consistent with the vertex of the cone.

\section{Discussion}

\subsection{Black hole mass estimates in the mid-infrared}

In order to prove the existence of a black hole or other compact object at the
center of NGC 7582 with spatially resolved gas dynamics, the resolution must 
be sufficient to probe gas dynamics inside the black hole's sphere of 
influence. The current data do not have the required spatial resolution for 
this, but if we accept that black hole activity is responsible for the AGN 
activity, the data 
presented here can be used to infer constraints on the mass of the 
black hole. We find a best-fitting black hole mass of 5.5$\times$10$^{7}$ 
M$_{\odot}$, with a 95\% confidence interval of [3.6,8.1]$\times$10$^{7}$ 
M$_{\odot}$, consistent with the prediction of 
$\approx 4\times10^{7}$ M$_{\odot}$ from the local $M_{\rm BH}$--$\sigma_{*}$ 
relation.

This is the first time that a black hole mass in a galaxy except our own 
Milky Way system has been estimated from gas 
dynamics using line emission in the mid-infrared. Usually H$\alpha$ and/or 
[\ion{N}{ii}] in the optical have been used, in particular for HST data.
Ground-based work has so far been limited to, and has relied on, emission 
lines in the near-infrared since better resolution is achieved at longer 
wavelengths. More measurements are expected to appear in the near-infrared
with the application of adaptive optics \citep{haring-neumayer06}. 
The VISIR data presented here are diffraction limited and therefore have the 
best obtainable spatial resolution possible with an 8-m 
telescope at the position of the [\ion{Ne}{ii}] line. 
The black hole in NGC 7582 seems to be on the limit of what is possible to 
constrain with a spatial resolution of 0\farcs4, and probably 
black hole masses of $\ga 5$--$10\times10^{7}$ M$_{\odot}$ in nearby galaxies 
with gas disks can be estimated from the ground using spectroscopy of the 
[\ion{Ne}{ii}]12.8$\mu$m line.

Since dust is easily penetrated at mid-infrared wavelengths, 
gas dynamics in the mid-infrared may prove to be a valuable tool for probing 
the $M_{\rm BH}$--$\sigma_{*}$ relation in galaxies where the nuclear regions 
are obscured
by dust and where optical and/or near-infrared spectroscopy may fall short.
Gas dynamics in the mid-infrared may therefore help determine the black hole 
masses of e.g.\ Seyfert~2 galaxies where the nucleus is hidden behind an
optically thick, dusty torus. The assumption is, as with optical and 
near-infrared observations, that the gas is well-behaved and has a relatively 
smooth rotation curve.

\subsection{The morphology of the circum-nuclear disk}

The two bright regions in the disk to the South of the nucleus may be dense 
regions of intense starformation, deeply embedded in dust \citep{wg06}. The 
two knots are equally bright in [\ion{Ne}{ii}], but one of them is less bright,
or undetected, in H$_{2}$ and at 11.9$\mu$m \citep{sosabrito01,siebenmorgen04}.
Our modeling shows that the rotation curve of the disk, including the two 
bright knots to the South of the nucleus, is consistent with a thin disk 
rotating in the gravitational potential of the bulge. This suggests that the 
two knots lie in the plane of the disk. According to the unified model the 
ionization cone axis is perpendicular to the plane of the torus, and since the
circum-nuclear disk is probably an extension of the torus, the two knots must 
be shielded from the ionizing continuum of the AGN. We therefore conclude that
the AGN continuum cannot be responsible for the destruction of PAHs and 
H$_{2}$ in one of the knots. The lack of PAHs and H$_{2}$ in one knot relative 
to the other may be explained by differences in the ionizing continuum and/or 
by differences in the physical conditions in the molecular clouds. For 
instance, the PAH- and H$_{2}$-deficient knot may be further along in its 
lifetime and has ionized more of its local interstellar medium, destroying 
both PAHs and H$_{2}$ in the process. Alternatively, the PAH-deficient knot 
may contain a larger population of hot O-stars than the other since PAHs are 
found to be better tracers of B-star populations rather than very hot O-stars 
\citep{peeters04}. Finally, as PAHs and H$_{2}$ can arise in different regions
and H$_{2}$ is very temperature-sensitive, differences in the physical 
conditions of the two knots may also influence their appearances.

\section{Conclusions}

We have presented data taken with the ESO VLT mid-infrared 
instrument, VISIR, consisting of a narrow-band image centered on the 
[\ion{Ne}{ii}]12.8$\mu$m line, and two high-resolution
($R = 16000$) spectra also centered on the [\ion{Ne}{ii}] line, of the
composite starburst-Seyfert~2 galaxy NGC 7582.
The galaxy has a circum-nuclear star forming disk and the 
two long-slit spectra were obtained with the slit parallel and perpendicular 
to the major axis of the disk. The parallel spectrum shows a clear rotation 
curve, with an amplitude of $\approx 190$ km\,s$^{-1}$, and the perpendicular 
spectrum, a roughly constant velocity profile as expected for a geometrically 
thin disk.   

We model the disk as a thin circular disk rotating in the combined 
gravitational potential of the bulge and a central black hole, and find a 
best-fit black hole mass of 5.5$\times$10$^{7}$ M$_{\odot}$ with a 95\% 
confidence interval of [3.6,8.1]$\times$10$^{7}$ M$_{\odot}$. In the modeling 
we also allow for the fact that the dynamical center may not have been exactly 
at the center of the slit during observation. The best-fitting model indicates
that it probably was located
close to the edge of the slit, approximately 0\farcs3 to the west of the slit 
center. Taking advantage of the high spectral resolution, we argue that the 
(heliocentric) systemic velocity of the galaxy is $cz \approx 1614$--1634 
km\,s$^{-1}$. 

There are two dense starforming knots in the disk, equally bright in 
[\ion{Ne}{ii}], but one shows weaker emission in H$_{2}$ and at 11.9$\mu$m, 
indicating that the PAHs and the H$_{2}$ may have been destroyed. Since the 
rotation in the outer parts of the disk, including the two knots, is well 
explained by a thin disk rotating in the gravitational potential of the bulge,
we conclude that the two knots lie in the disk and are thereby shielded from
the AGN continuum. If an ionizing continuum is responsible for the 
destruction of PAHs and H$_{2}$ in one of the knots, it must come from the 
star forming region itself rather than the AGN. It cannot be excluded however,
that different physical conditions, e.g.\ temperature, within the two knots 
may be responsible for their different appearances in H$_{2}$ and at 
11.9$\mu$m.

The data are diffraction limited with a spatial resolution of 
$\approx 0\farcs4$, 
and we argue that this resolution is sufficient to put constraints on 
black holes masses of $\ga 5$--$10\times10^{7}$ M$_{\odot}$. 
Now that STIS on board HST is no longer operational, other techniques for 
estimating black hole masses in galaxies need to be explored. We have shown 
that gas dynamical estimates of black hole masses based on high-resolution 
mid-infrared spectroscopy are competitive with other techniques, particularly
if the galaxy nucleus is obscured by dust. Such observations also allow us 
to study obscured star formation in the centers of nearby AGN, thus advancing 
our understanding of the connection between AGN activity and nuclear star 
formation.

\begin{acknowledgements}
We are grateful to the VISIR Science Verification team at Paranal for 
performing the observations. M. Cappellari is acknowledged for making the  
MGE\_FIT\_SECTORS package publically available, and G. Verdoes-Kleijn
and E. Galliano are thanked for discussions. The referee is acknowledged
for comments which helped improve the manuscript. NSO/Kitt Peak FTS data used
here were produced by NSF/NOAO. This research has made use of 
the NASA/IPAC Extragalactic Database (NED) which is operated by the Jet
Propulsion Laboratory, California Institute of Technology, under contract
with the National Aeronautics and Space Administration.

\end{acknowledgements}

\bibliographystyle{aa}
\bibliography{/Users/wold/publications/tex/references}

\end{document}